\documentclass[10pt,a4paper]{article}
\usepackage[margin=0.5 in]{geometry}
\usepackage{cite}
\usepackage{amsmath,amssymb,amsfonts}
\usepackage{lipsum} 
\usepackage{graphicx}
\usepackage{textcomp}
\usepackage{multirow}
\usepackage{subcaption}
\usepackage{longtable}
\usepackage{xcolor}
\usepackage{mathtools}
\usepackage{breqn}
\usepackage{comment}
\usepackage{rotating}
\usepackage{longtable}
\usepackage{makecell}
\usepackage{flushend}
\usepackage{makecell}
\PassOptionsToPackage{hyphens}{url}
\usepackage{hyperref}
\usepackage{url}
\usepackage{tablefootnote}
\usepackage{color}
\usepackage[normalem]{ulem}

\providecommand{\keywords}[1]
{
	\small	
	\textbf{\textit{Keywords---}} #1
}
\def\BibTeX{{\rm B\kern-.05em{\sc i\kern-.025em b}\kern-.08em
		T\kern-.1667em\lower.7ex\hbox{E}\kern-.125emX}}
\begin{document}
	
	\title{ Synthesis of Ventilator Dyssynchrony Waveforms using a Hybrid Generative Model and a Lung Model\\
	}
	\date{}
	\author{\tmp}
	\author{
		\normalsize Sagar Deep Deb$^{1}$, Suvakash Dey$^{1}$, Deepak K. Agrawal$^{1*}$ \\
		\small $^{1}$Department of Biosciences and Bioengineering, Indian Institute of Technology Bombay, Mumbai, India\\
		\small *Email: \texttt{dkagrawal@iitb.ac.in}
	}


	\maketitle
\begin{abstract}
Ventilator dyssynchrony (VD) is often described as a mismatch between a patient breathing effort and the ventilator support during mechanical ventilation. This mismatch is often associated with an increased risk of lung injury and longer hospital stays. The manual VD detection method is unreliable and requires considerable effort from medical professionals. Automating this process requires a computational pipeline that can identify VD breaths from continuous waveform signals. For that, while various machine learning (ML) models have been proposed, their accuracy is often limited due to the unavailability of a large, well-annotated VD waveform dataset. This paper presents a new approach combining mathematical and deep generative models to generate synthetic, clinically relevant VD waveforms. The mathematical model, which we call the VD lung ventilator model (VDLV), can accurately replicate clinically observable deformation in the pressure and volume waveforms. These temporal deformations are hypothesized to be related to specific VD breaths. We leverage the VDLV model to produce training waveform datasets covering normal and various VD breaths. These datasets are further diversified using deep learning models such as Generative Adversarial Network (GAN) and Conditional GAN (cGAN). The performance of both GAN and cGAN models is assessed through quantitative metrics, demonstrating that this hybrid approach effectively creates realistic and diverse VD waveforms. Notably, the pressure and volume cGAN models enable the generation of more precise and targeted VD signals. These improved synthetic waveform datasets have the potential to significantly enhance the accuracy and robustness of VD detection algorithms.		
\end{abstract}
	

\section{Introduction}
Mechanical ventilation (MV) is a life-saving measure for patients who are unable to achieve the necessary gas exchange. However, improper MV can lead to ventilator-induced lung injury (VILI) \cite{de2021patient}. One major contributor to VILI is ventilator dyssynchrony (VD), which occurs when ventilator settings are inadequate and fail to match the patient respiratory demands. VD is often associated with high mortality, where different types of VD potentially cause varying degrees of damage \cite{gao2022early, sottile2020ventilator}. The interplay between VD and VILI is often challenging to understand due to the complex dynamics of ventilator mechanics, rapidly evolving pulmonary physiology, and various healthcare interventions. As a result, the specific pathways via which VD can damage lungs are not well understood \cite{oto2021patient,de2019patient,de2021patient}.
\\
In current clinical practice, identifying VD often involves manually analysing pressure-time, volume-time, and flow-time waveforms on ventilators. This method is labor-intensive and susceptible to errors due to inter-observer variability \cite{pan2021identifying}. Recent efforts to automate the VD detection process employ ML and deep learning (DL) models. These approaches typically fall into two categories: 1) classifying VD waveforms directly using time series data classification methods \cite{sottile2023development, zhang2020detection, pan2021interpretable, bakkes2023automated}, and 2) transforming the time series data into images for subsequent classification tasks \cite{pan2021identifying, obeso2023novel}. For example, Zhang et al. \cite{zhang2020detection} implemented a two-layer long short-term memory (LSTM) network for detecting double trigger and ineffective trigger VD. Following this, an interpretable 1-D CNN architecture supplemented with an LSTM model was developed to classify various VD types, such as double trigger, delayed cycling, early cycling, and ineffective trigger VD \cite{pan2021interpretable}. Bakkes et al. \cite{bakkes2023automated} introduced a U-Net-based model for identifying the start and end of inspiration, utilizing esophageal pressure measurements for VD classification through simulated data \cite{van2022model}. 
In a recent study, Sottile et al. \cite{sottile2023development} used manual feature engineering to identify seven types of VD using features like inspiratory time, expiratory time, peak width, and peak counts. A training dataset was created by manually labeling 10,000 breaths from 30 ICU patients and used to train four machine learning models, with XGBoost having the highest specificity. On the other hand, Pan et al. \cite{pan2021identifying} utilized transfer learning \cite{zhuang2020comprehensive} to transform 1-D ventilator signals into 2-D images for classifying double trigger and ineffective trigger VD. Similarly, Obeso et al. \cite{obeso2023novel} used spectrograms of pressure and flow signals to train a VGG-Net and to classify waveforms into two different categories: reverse trigger and inadequate support. The inadequate support combined with early cycling and flow limited VD. While these approaches are promising, they often rely on creating a hand-annotated training dataset and manual feature extraction, which limit the overall performance and effectiveness of ML and DL models
\\
A well-annotated, balanced dataset is important for training ML and DL models to generalize to unseen data \cite{ahmed2023deep}. However, creating such datasets is challenging due to the need for manual annotation by medical professionals \cite{sottile2020ventilator, pan2021identifying, bakkes2023automated}. Moreover, waveform data collection in ICU settings can be challenging due to logistical and ethical issues\cite{karahoda2024generating,rai2024gan}. To overcome these challenges, researchers employed Generative Adversarial Networks (GANs) \cite{goodfellow2020generative} and conditional GAN (cGAN) \cite{mirza2014conditional} for generating complex biomedical data such as electrocardiograms \cite{karahoda2024generating,rai2024gan} and medical images \cite{lee2021identifying,showrov2024generative}. Using a similar approach, Hao et al. \cite{hao2024adversarial} developed a generative VentGAN model to produce synthetic VD waveforms. The model is trained on synthetic waveform datasets produced using a non-linear lung-airway mathematical model \cite{van2022model}. However, due to the limited diversity of waveforms in the initial training dataset, the model is limited to trigger and cycling-based dyssynchrony. As a result, the full range of VD variability has not yet been explored via GAN models.
\\
Our long-term goal is to create a computational pipeline for automating VD detection while reducing the need for manual feature extraction and annotation. To achieve this, a diverse set of VD waveforms that effectively capture the variability inherent in ventilator waveforms is required. This study presents a synthetic ventilator waveform generator that produces clinically relevant pressure and volume waveforms using a combination of mathematical modeling and a deep generative network. The underlying mathematical model can produce normal and VD associated pressure and volume waveforms with adjustable temporal features that are pathophysiological relevant~\cite{agrawal2024quantifiable, agrawal2021damaged}. Here, we expand this model to produce six types of VD waveforms, and refer to it as the VD lung ventilator (VDLV) model. By employing the VDLV model, we develop training datasets for five relevant VD types, including normal waveforms. These datasets are then used to train GAN and cGAN models, which further enhance the diversity present in the waveforms.
\\
The paper is organized as follows: Section \ref{second_section} describes the VD-related deformation in the pressure and volume waveforms, while Section \ref{association} links these deformations with the mechanical ventilator modes and parameters. Sections \ref{MathematicalModeling} and \ref{section5} explain the formulation of the VDLV model and its application in producing various VD waveforms. Section \ref{HeterogeneityusingGAN} explains the use of GANs and cGANs to improve the diversity of waveforms. Section \ref{results_and_discussions} presents the qualitative and quantitative evaluation of the ventilator waveforms with respect to GANs and cGAN models along with limitations and suggests directions for future research. Finally, Section \ref{conclusion} summarizes the key findings.
	
\section{Identifying key deformations in VD waveforms } 
\label{second_section}
	
\begin{figure*}[h]
		\centering
		\hspace{-2cm}
		\subfloat[]{\includegraphics[scale=0.080]{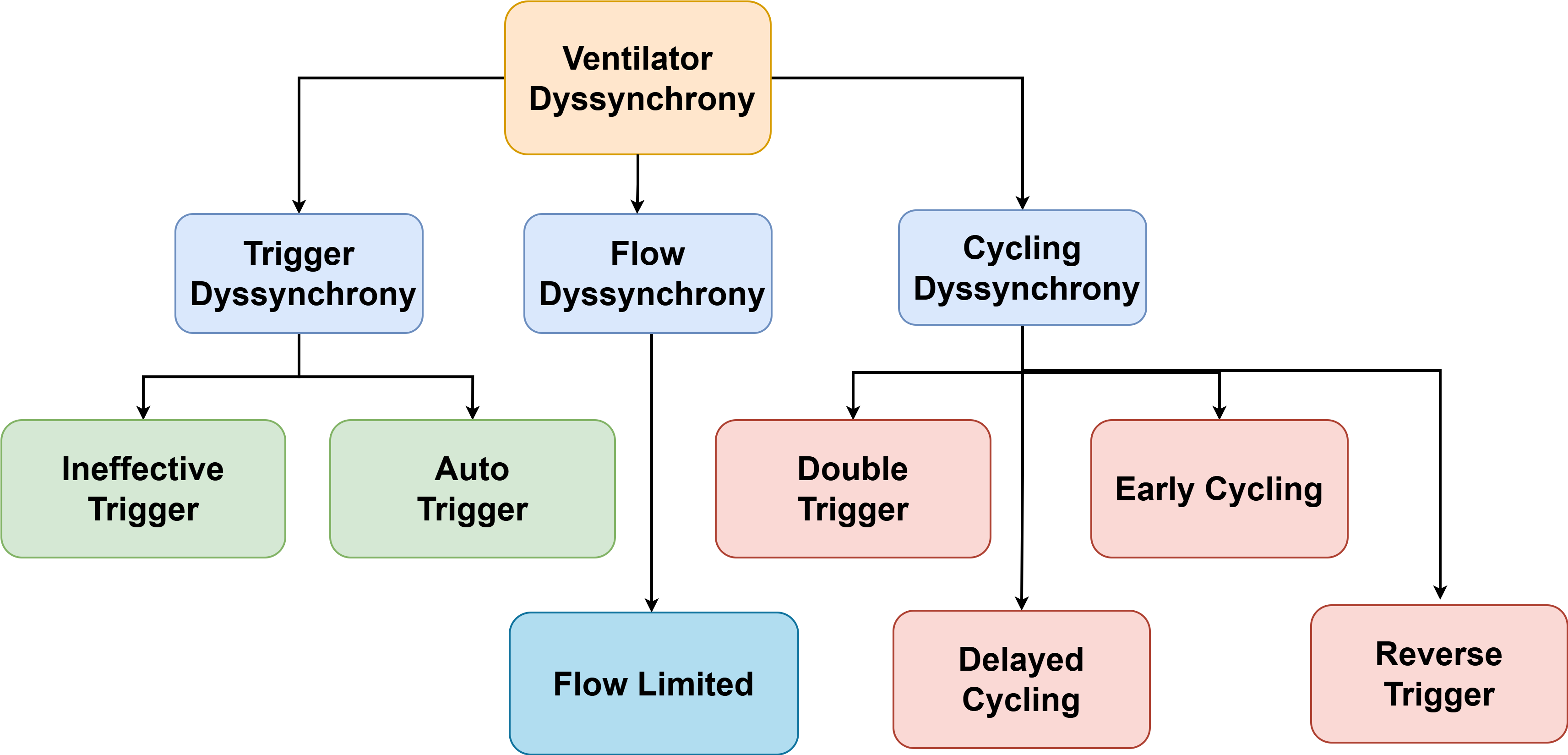}%
			\label{block_Dia}} 
		\hspace{1cm}
		\subfloat[]{\includegraphics[scale=0.4]{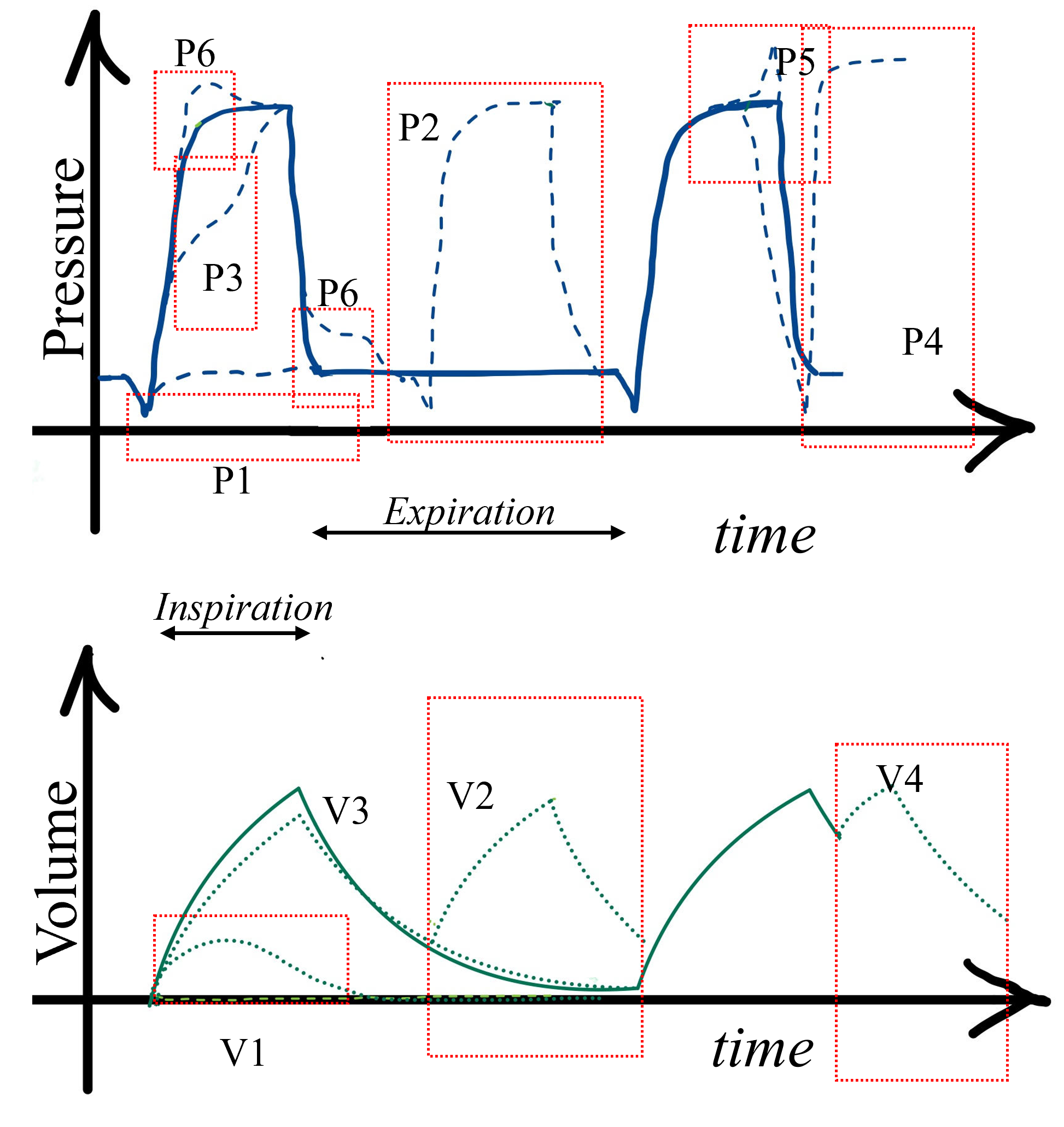}%
			\label{example}}
		\caption{(a) The flowchart showing various types of ventilator dyssynchrony (VD) (b) Pictorial representation of VD associated distinct time-dependent deformation in the pressure and volume waveforms. }
\end{figure*}

Based on the literature, we identify seven clinically relevant VD, which are broadly categorized into three main groups: trigger, flow, and cycling dyssynchrony (see Fig. \ref{block_Dia}) \cite{oto2021patient}. Trigger dyssynchrony occurs when there is a mismatch between a patient respiratory effort and the ventilator response \cite{de2011monitoring}. Flow dyssynchrony arises when the ventilator airflow fails to adequately satisfy the patient inspiratory demands \cite{oto2021patient}. Cycling dyssynchrony occurs when the timing set by the ventilator for transitioning from inhalation to exhalation is mismatched with the patient desire to end inhalation \cite{de2021patient}. These main dyssynchrony groups are divided into seven specific VD subtypes, as shown insee Fig. \ref{block_Dia}. Each VD produces distinct time-dependent deformations in the pressure, volume, and flow waveforms, which are often used by clinicians to understand the nature of the patient-ventilator interaction \cite{mellott2009patient}. Since the flow signal can be derived from the volume signal, we do not analyse the flow signal to identify the key VD-related deformations. 
	
Figure~\ref{example} illustrates several VD-related deformations in the pressure and volume waveforms, labeled P and V, respectively. Features P1 and V1 indicate an ineffective trigger VD, where the ventilator fails to deliver a breath in response to the patient effort \cite{de2011monitoring}. In contrast, P2 and V2 represent auto trigger VD, where the ventilator delivers a breath in the absence of patient demand \cite{de2011monitoring}. Feature P3 highlights flow limited (also known as flow starvation) VD, where the ventilator fails to meet the patient demand for inspiratory flow \cite{mellott2009patient}. This flow limitation is noticeable in the volume waveform with a low inspiratory rise time (Feature V3). 
Features P4 and V4 illustrate double trigger VD in which two breaths are delivered without sufficient time to exhale during a sustained patient effort \cite{mellott2009patient}. Feature P5 demonstrates delayed cycling VD, which is characterized by the prolonged inspiration that extends beyond the patient inspiratory time \cite{oto2021patient}. Feature P6 shows early cycling VD (also known as premature cycling), where expiration is initiated before the patient has completed inhalation \cite{oto2021patient}. Collectively, delayed and early cycling VD primarily influence the timing of the transition from inhalation to exhalation in the pressure waveforms without producing significant observable changes in the volume waveforms \cite{sottile2020ventilator,oto2021patient,de2011monitoring}. Finally, reverse trigger VD, which may be considered a variant of double trigger VD, occurs when passive ventilator breaths trigger a neural response. This can result in involuntary diaphragmatic contractions \cite{akoumianaki2013mechanical, de2021patient, murray2022reverse}. Since the flow signal can be derived from the volume signal, we do not analyse the flow signal to identify the key VD-related deformations.

\section{Associating occurrence of VD with Ventilator Parameters and Modes} 
	\label{association}

\begin{figure*}[!h]
	\centering
	\begin{subfigure}[b]{0.3\textwidth}
		\centering
		\includegraphics[height=4.5cm]{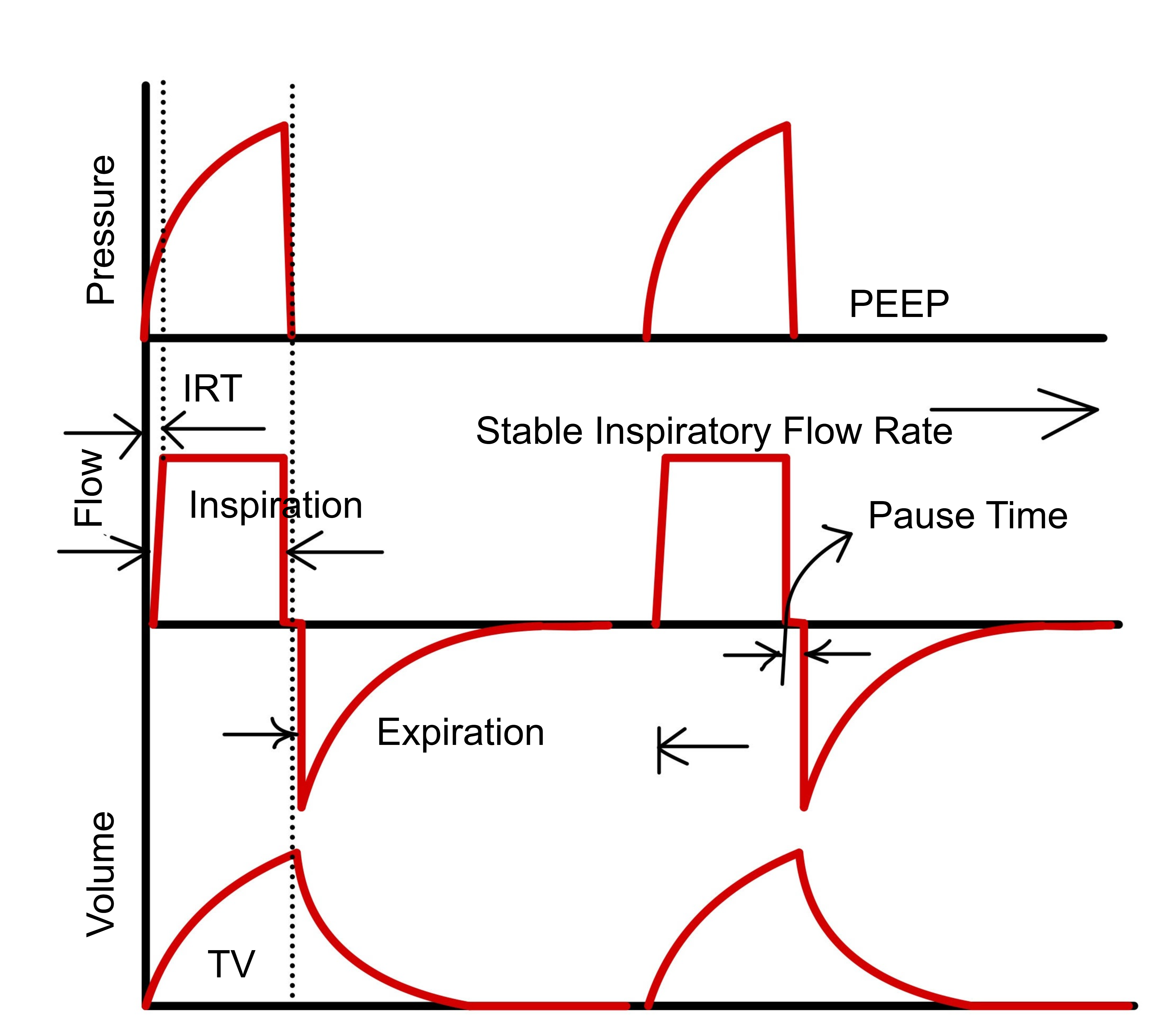}
		\caption{}
		\label{fig:VCV}
	\end{subfigure}
	\hfill
	\begin{subfigure}[b]{0.3\textwidth}
		\centering
		\includegraphics[height=4.5cm]{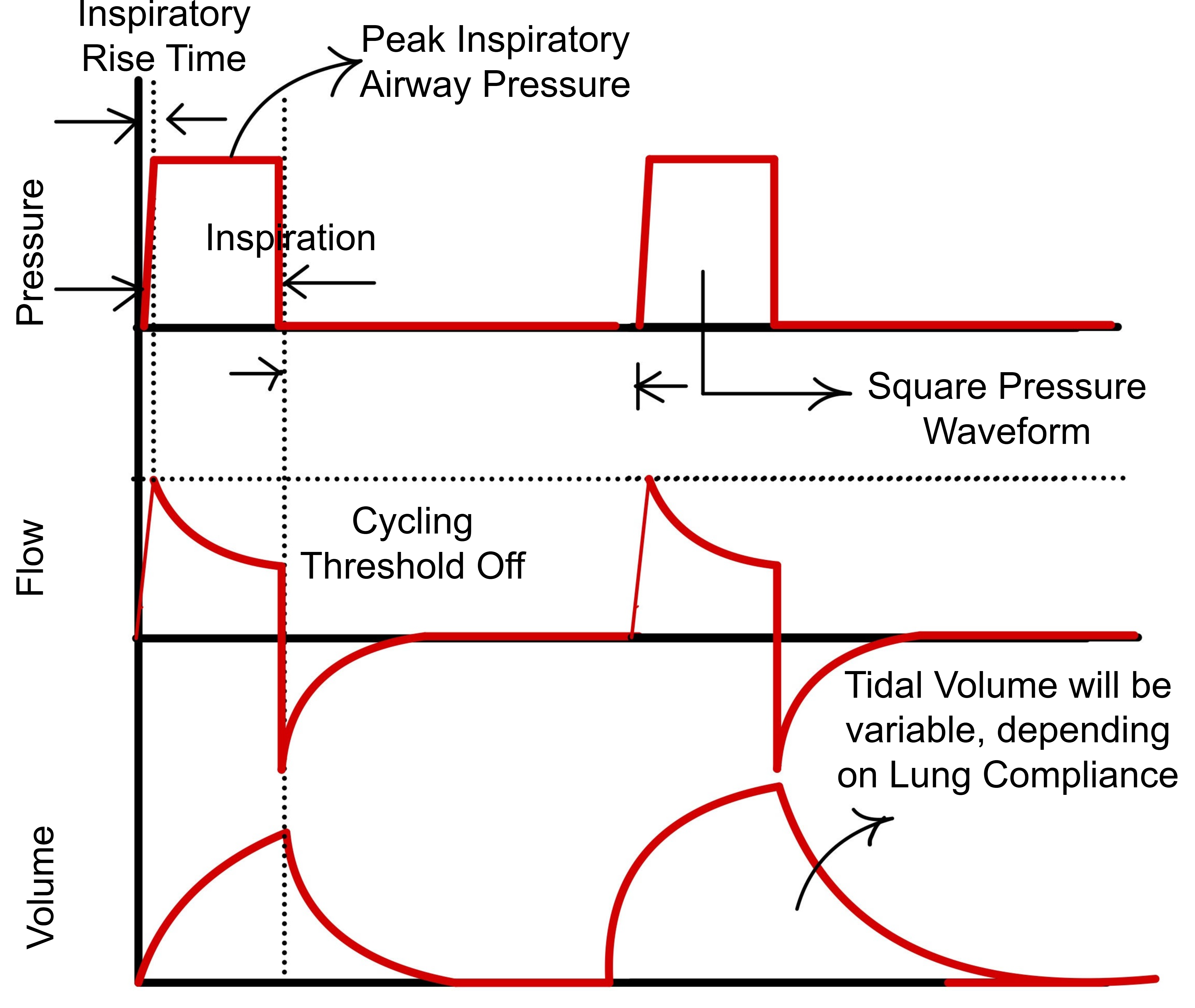}
		\caption{}
		\label{fig:PVC}
	\end{subfigure}
	\hfill
	\begin{subfigure}[b]{0.3\textwidth}
		\centering
		\includegraphics[height=4.5cm]{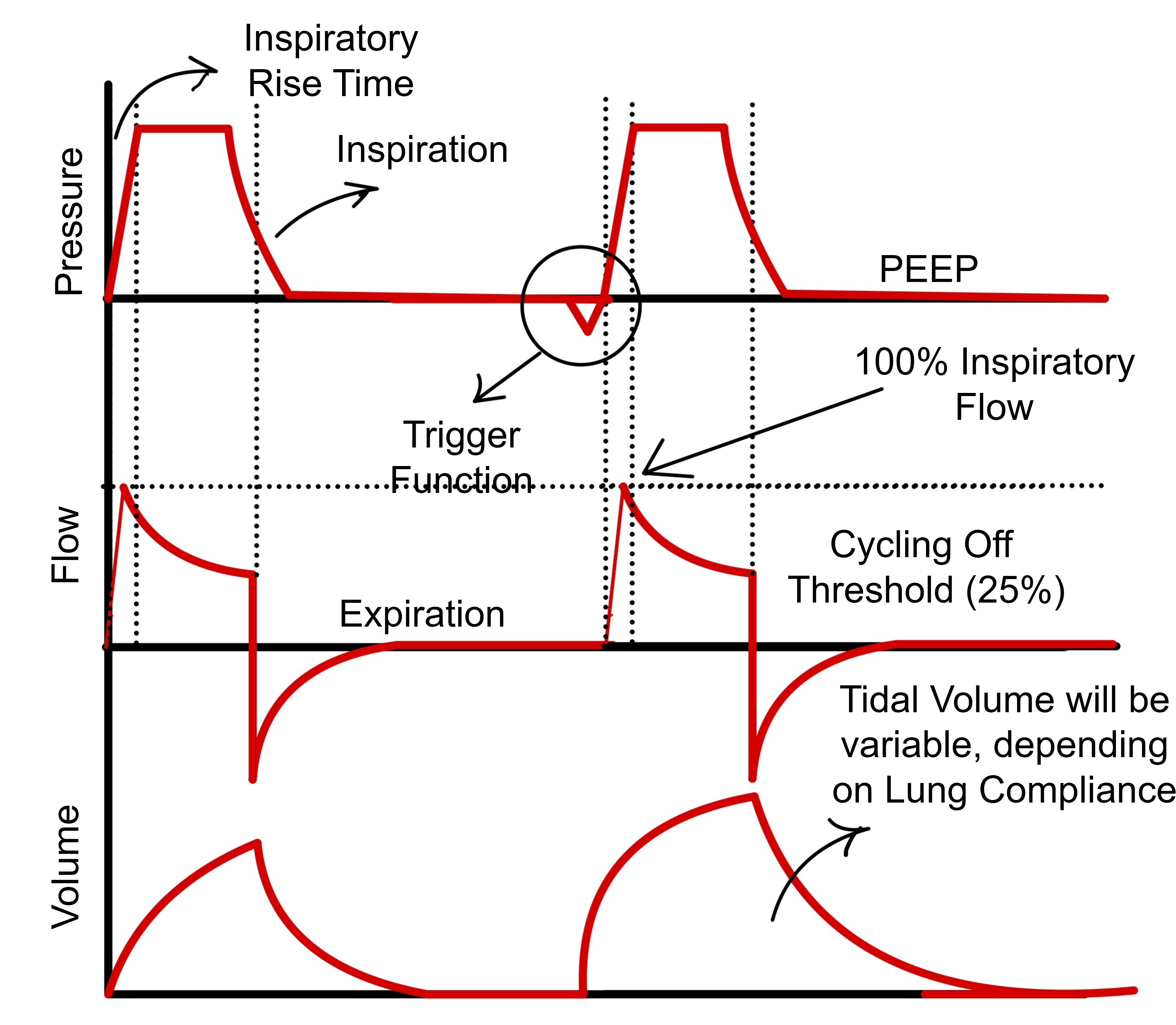}
		\caption{}
		\label{fig:PSV}
	\end{subfigure}
	\caption{
		Schematic representation of pressure, flow, and volume waveform signals in 
		(a) Volume Control Ventilation mode \textit{(TV, RR, PEEP, $FiO_2$, I:E, PT, IRT, TF)}, 
		(b) Pressure Control Ventilation mode \textit{(PC+, RR, PEEP, $FiO_2$, I:E, IRT, TF)}, and 
		(c) Pressure Support Ventilation mode \textit{(PS+, PEEP, $FiO_2$, IRT, ICO)} 
		\cite{brander2008invasive,chatburn2007classification}. 
		These modes are supported by the Servo-i ventilator.
	}
	\label{ventilator_modes}
\end{figure*}

The occurrence of VD varies significantly across different ventilator modes, as each mode operates under a specific set of ventilator parameters that control patient-ventilator interaction. Table \ref{table:Abbreviations} summarizes typical ventilator parameters while Fig. \ref{ventilator_modes} illustrates the relationships between these parameters and the waveforms. Given that the nomenclature for ventilator parameters can vary among manufacturers, we have compiled relevant information for the Maquet Servo-i ventilator \footnote{\url{https://www.medonegroup.com/pdf/manuals/userManuals/Maquet-Servo-i-Operators-Manual.pdf}}.

	\begin{table}[!h]
		\caption{Ventilation parameters for the Maquet Servo-i ventilator.}
		\label{table:Abbreviations}
		\centering
		\begin{tabular}{|cp{5cm}|p{9cm}|}
			\hline
			\multicolumn{2}{|c|}{\textbf{Abbreviations}} & \textbf{Remarks} \\ \hline
			\multicolumn{1}{|c|}{RR} & Respiratory Rate & Number of breaths delivered by the ventilator per minute. \\ \hline
			\multicolumn{1}{|c|}{TV} & Tidal Volume & Pre-set volume of air delivered to the patient during each breath. \\ \hline
			\multicolumn{1}{|c|}{ICO} & Inspiratory Cycle Off & Defines when the ventilator transitions from inspiration to expiration.\\ \hline
			\multicolumn{1}{|c|}{PEEP} & \begin{tabular}[c]{@{}c@{}}Positive End Expiratory Pressure\end{tabular} & Constant baseline pressure value at the end of expiration to prevent alveolar collapse. \\ \hline
			
			\multicolumn{1}{|c|}{PC+} & \begin{tabular}[c]{@{}c@{}}Pressure Control+ \end{tabular} & Ensures each breath achieves the target inspiratory pressure above PEEP. \\ \hline
			\multicolumn{1}{|c|}{PS+} & \begin{tabular}[c]{@{}c@{}}Pressure Support+ \end{tabular} & Provides additional pressure to support each patient-initiated breath above PEEP.\\ \hline
			
			\multicolumn{1}{|c|}{FiO2} & \begin{tabular}[c]{@{}c@{}}Fraction of Inspired Oxygen\end{tabular} & Percentage of oxygen delivered with each breath. \\ \hline
			\multicolumn{1}{|c|}{I:E} & \begin{tabular}[c]{@{}c@{}}Inspiration to Expiration Ratio\end{tabular} & Ratio between the duration of inhalation and exhalation during each breath.\\ \hline
			\multicolumn{1}{|c|}{PT} & Pause Time & Brief pause at the end of inhalation to improve gas exchange before exhalation begins. \\ \hline
			\multicolumn{1}{|c|}{IRT} & Inspiratory Rise Time & Time taken for the ventilator to reach the set inspiratory flow or pressure target.\\ \hline
			\multicolumn{1}{|c|}{TF} & Trigger Function & Sensitivity setting to detect patient efforts to start a breath. \\ \hline
		\end{tabular}
	\end{table}

These ventilator parameters are controlled by different modes, which are broadly categorized into control mode and support mode \cite{oto2021patient, mirabella2020patient}. In control mode, the ventilator autonomously manages breath delivery based on preset settings, while support mode assists the patient spontaneous breaths \cite{oto2021patient}. Based on this, the Maquet Servo-i ventilator primarily operates in three modes: Volume-Controlled Ventilation (VCV), Pressure-Controlled Ventilation (PCV), and Pressure Support Ventilation (PSV) (see Fig. \ref{ventilator_modes}). Table \ref{ventilator_modes} illustrates a relationship between VD and ventilator parameters with respect to each mode. 
	
	\begin{table*}[h!]
		\centering
		\caption{Different ventilator modes, dyssynchrony, and associated parameters. }
		\label{tab:table2}
		\begin{tabular}{|l| |c| |c| |c| }
			\hline
			\textbf{VD/modes} & \textbf{Volume control} & \textbf{Pressure control} & \textbf{Pressure support} \\\hline
			Ineffective Trigger   & TF, PEEP \cite{oto2021patient}  & TF \cite{oto2021patient}, PEEP \cite{gilstrap2013patient}, PC+ \cite{oto2021patient} & TF, PEEP \cite{oto2021patient}, PS+ \\ \hline
			Auto Trigger     & TF   & TF \cite{oto2021patient} & TF  \\ \hline
			Flow Limited      & TV, IRT \cite{mellott2009patient}     & IRT \cite{oto2021patient,mellott2009patient}   & IRT\cite{oto2021patient,mellott2009patient}   \\ \hline
			
			Double Trigger     & I:E \cite{antonogiannaki2017patient}, TV \cite{oto2021patient}    & I:E \cite{antonogiannaki2017patient}, PC+    & ICO, PS+ \cite{oto2021patient,vignaux2009patient}  \\ \hline		
			Delayed Cycling     & I:E, TV \cite{de2011monitoring}   & I:E \cite{oto2021patient}           & ICO \cite{antonogiannaki2017patient}       \\ \hline
			Early Cycling     & I:E,TV \cite{de2011monitoring}    & I:E \cite{oto2021patient}  & ICO \cite{antonogiannaki2017patient} \\ \hline
			
		\end{tabular}
	\end{table*}

In trigger dyssynchrony, the TF parameter plays an important role in determining the delivery of a ventilator breath. Ineffective trigger VD typically occurs when the effort made by the patient fails to reach the trigger threshold, which can be pressure or flow, set in the ventilator using the TF parameter \cite{gilstrap2013patient}. This is one of the most common forms of VD \cite{mellott2009patient,sottile2020ventilator}, which is found in all three ventilator modes as shown in Table \ref{tab:table2}. Ineffective trigger can also occur due to incomplete exhalation, resulting in a rise in the baseline pressure (also known as auto-PEEP) \cite{oto2021patient}. In PCV and PSV modes, PC+ and PS+ settings can also increase baseline pressure. Auto trigger VD can also occur in any of the three modes when a breath is delivered by the ventilator, even in the absence of a patient effort, which is generally due to the low TF values \cite{vignaux2009patient,de2021patient, oto2021patient }. Moreover, during MV, flow delivery can be either fixed, as in VCV, or dynamically adjusted to the patient inspiratory effort, as in PCV and PSV \cite{oto2021patient}. In VCV, if the preset flow rate, which is controlled by the TV and I:E parameters, fails to meet the patient inspiratory demand, flow limited VD may occur \cite{oto2021patient, de2021patient} (see Table \ref{tab:table2}). In PCV and PSV, inappropriate IRT can also cause flow limited VD due to variable flow rates \cite{de2011monitoring,mellott2009patient}. 
\\
Double trigger VD occurs when the patient neural inspiratory time exceeds the inspiratory time set by the ventilator \cite{liao2011classifying}. The inspiratory time is regulated by the I:E and ICO parameters in control mode and support mode, respectively \cite{oto2021patient} (Table \ref{tab:table2}). This VD can also occur when the ventilator does not adequately meet the patient inspiratory effort. Therefore, the TV parameter in VCV, PC+ in PCV, and PS+ in PSV are also often related to the occurrence of double trigger VD \cite{vignaux2009patient,antonogiannaki2017patient}. Finally, delayed and early cycling VD occur when ventilator timing does not align with the patient respiratory cycle \cite{oto2021patient}. Delayed cycling VD happens when the ventilator inspiratory time exceeds the patient neural inspiratory duration, leading to premature termination of inspiration. Conversely, early cycling VD occurs when the ventilator switches from inspiration to expiration while the patient continues in the inspiration phase. Unlike double trigger VD, where a strong patient effort unintentionally triggers a second breath \cite{daniel2017identifying,sottile2020ventilator,gilstrap2013patient}, in early cycling, VD involves premature cycling during inspiration. In VCV, inappropriate I:E and TV settings can contribute to early and delayed cycling VD. Similarly, in PCV and PSV, incorrect I:E and ICO parameters contribute to these VD, respectively (see Table \ref{tab:table2}). 
	
\section{Formulation of VD-lung ventilator model}
\label{MathematicalModeling}
	
The basic framework of the VDLV model remains the same, and a detailed explanation can be found at \cite{agrawal2024quantifiable, agrawal2021damaged}. Here, we further modify this model to produce clinically relevant VD waveform signals \cite{simpson2024making}. The VDLV model uses a periodic rectangular waveform to which functional terms are added to produce specific features in the pressure and volume waveforms. Deformations in these features may reflect time-sensitive lung conditions and patient-ventilator interactions \cite{agrawal2021damaged}. The pressure waveform model ($P$) is defined as:
	
	\begin{equation}
		P (t,\theta) = f_{p1}(t,\theta) + f_{p2}(t,\theta) + f_{p3}(t,\theta) + CP.
		\label{pressure_equation}
	\end{equation}
	
	Here, the $f_{p1}$ component is the baseline signal, producing an idealized pressure signal. The $ f_{p2}$ component introduces deformation at the beginning or during inspiration, while the $ f_{p3}$ component can be used to produce the deformation at the end of inspiration or during exhalation. These deformations represent dyssynchronous patient-ventilator interaction. The CP parameter is constant baseline pressure, which is referred to as PEEP \cite{agrawal2024quantifiable}. The $f_{p1}$ component is defined as:
	
	\begin{equation}
		f_{p1}(t,\theta) = A_{p1}\{f_{p11}(t,\theta) + f_{p12}(t,\theta)\},
	\end{equation}
	where,
	
	\begin{equation}
		\begin{split}
			f_{p11} (i+1) = \{\frac{1}{\gamma_{p1}} f_{pb1}(i+1) + (1-\frac{1}{\gamma_{p1}}) f_{p11}(i)\}; i=1:n, \\
			f_{p11}(t,\theta) = [f_{p11}(1) f_{p11}(2)...f_{p11}(i)...f_{p11}(n)] \frac{f_{pb1}}{\max B[f_{p11}(t,\theta)]},
		\end{split}
	\end{equation}
	and,
	\begin{equation}
		\begin{split}
			f_{p12} (i+1) = \{\frac{1}{\gamma_{p2}} f_{pb1}(i+1) + (1-\frac{1}{\gamma_{p2}}) f_{p12}(i)\}; i=1:n,\\
			f_{p12}(t,\theta) = [f_{p12}(1) f_{p12}(2)...f_{p12}(i)...f_{p12}(n)] \frac{(1-f_{pb1})}{\max B[f_{p12}(t,\theta)]}.
		\end{split}
	\end{equation}
	
	The $f_{p11}$ and $f_{p12}$ sub-components produce the inspiration and the expiration part of the pressure signal, respectively, with $f_{p11} (i=1)= f_{p12} (i=1) = 0$. Here, $\max B $ corresponds to the maximum value of the respective component for the entire breath. The parameter $\gamma_{p}$ with the numerical subscript allows to alter the gradient of the respective signal. Additional deformations in the waveform signal are produced using the $f_{p2}$ and $f_{p3}$ components, respectively, which are given as:
	
	\begin{equation}
		f_{p2}(t,\theta) = A_{p2} \frac{f_{pb2}(t,\theta)}{\max B [f_{pb2}(t,\theta)]},
	\end{equation}

	\begin{equation}
		f_{p3} (t,\theta) = A_{p3} \frac{f_{pb3}(t,\theta)}{\max B [f_{pb3}(t,\theta)]}. 
	\end{equation}
	The parameter $A_{p}$ with numerical subscripts controls the amplitude of the individual components. The $f_{p1}$, $f_{p2}$, and $f_{p3}$ components use respective periodic signals $f_{pb1}$, $f_{pb2}$, and $f_{pb3}$ and are represented as
	\begin{equation}
		f_{pb1}(t,\theta) = \frac{1}{2}\{tanh(\alpha_{p1}(sin(2\pi\theta t - \phi_{p1})-\beta_{p1}))+1\},
	\end{equation}
	
	\begin{equation}
		f_{pb2}(t,\theta) = \frac{1}{2}\{tanh(\alpha_{p2}(sin(2\pi\theta t - \phi_{p2})-\beta_{p2}))+1\},
	\end{equation}
	
	and
	
	\begin{equation}
		f_{pb3}(t,\theta) = \frac{1}{2}\{tanh(\alpha_{p3}(sin(2\pi\theta t - \phi_{p3})-\beta_{p3}))+1\}.
	\end{equation}
	Here, parameters $\alpha_{p}$, $\phi_{p}$, and $\beta_{p}$ with respective subscripts control the smoothness, starting point, and duty cycle of the respective rectangular waveform. The volume model follows a similar structure and is defined as
	
	\begin{equation}
		V = f_{v1}(t, \theta) + f_{v2}(t, \theta)
		\label{volume_equation}.
	\end{equation}
	Here, the $f_{v1}(t,\theta)$ component produces the baseline volume signal, and is defined as:
	\begin{equation}
		f_{v1}(t,\theta) = A_{v1} \left( f_{v11}(t, \theta) + f_{v12}(t, \theta) \right),
	\end{equation}
	where the $f_{v11}$ and $f_{v12}$ sub-components produce the inspiration and the expiration part of the volume signal, respectively, with $f_{v11} (i=1)= f_{v12} (i=1) = 0$, which are given as:
	\begin{equation}
		\begin{split}
			f_{v11}(i+1) = \{\frac{1}{\gamma_{v1}} f_{vb1}(i+1) + \left(1 - \frac{1}{\gamma_{v1}}\right) f_{v11}(i)\}; i = 1:n, \\
			f_{v11} (t,\theta) = \left[ f_{v11}(1) \quad f_{v11}(2) \dots f_{v11}(i) \dots f_{v11} (n) \right] \frac{f_{vb1} (t,\theta)}{\max B \left[ f_{v11} (t,\theta) \right]},
		\end{split}
	\end{equation}
	and
	\begin{equation}
		\begin{split}
			f_{v12}(i+1) = \{\frac{1}{\gamma_{v2}} f_{vb1}(i+1) + \left(1 - \frac{1}{\gamma_{v2}}\right) f_{v12}(i)\}; i = 1:n, \\
			f_{v12} (t,\theta) = A_{v1} \left[ f_{v12}(1) \quad f_{v12}(2) \dots f_{v12}(i) \dots f_{v12} (n) \right] \frac{f_{vb1} (t,\theta)}{\max B \left[ f_{v12} (t,\theta) \right]}.
		\end{split}
	\end{equation}
	
	The component $f_{v2}(t,\theta)$ produces an additional deformation that can be used to reproduce features related to dyssynchrony in the volume signal and is expressed as:
	\begin{equation}
		f_{v2}(t, \theta) = A_{v2} \frac{f_{vb2}(t, \theta)}{\max B [f_{vb2}(t, \theta)]}.
	\end{equation}
	
	Both $f_{v1}$ and $f_{v2}$ components uses periodic rectangular waveform signal, defined as: 
	
	\begin{equation}
		f_{vb1}(t, \theta) = \frac{1}{2} \left\{ \tanh\left(\alpha_{v1} \left(\sin(2\pi\theta t - \phi_{v1}) - \beta_{v1} \right)\right) + 1 \right\},
	\end{equation}
	
	\begin{equation}
		f_{vb2}(t, \theta) = \frac{1}{2} \left\{ \tanh\left(\alpha_{v2} \left(\sin(2\pi\theta t - \phi_{v2}) - \beta_{v2} \right)\right) + 1 \right\}.
	\end{equation}
	
Parameters $\alpha_{v}$, $\phi_{v}$, and $\beta_{v}$ with subscripts control the smoothness, starting point, and duty cycle of the respective rectangular waveform. Additional description of the model parameters can be found at \cite{agrawal2024quantifiable}. 
	
The parameters of the VDLV model can be directly related to the ventilator parameters listed in Table \ref{table:Abbreviations}. Specifically, the model parameter CP corresponds to the PEEP, while the peak inspiratory pressure (PIP) and tidal volume (TV) are represented by $A_{p1}$ and $A_{v1}$, respectively. The inspiration-to-expiration (I:E) ratio in the control mode is altered by tuning the parameters $\beta_{p1}$ and $\beta_{v1}$ for the pressure and volume waveforms, respectively. In the support mode, the ICO parameter determines the cycling-off point and is similarly governed by the same set of parameters. The respiratory rate (RR) is parameterized using $\theta$ for both the pressure and volume waveforms, whereas the inspiratory rise time (IRT) is controlled through the parameter $\gamma_{v1}$ for the volume signal. Finally, we do not explicitly model the trigger factor (TF) parameter, which controls the initiation of a breath. However, the $\phi_{p1}$ and $\phi_{v1}$ parameters can be used to alter the onset of inspiration in pressure and volume waveforms, respectively.

\label{heterogenity}
\begin{figure*}
	\centering
	\includegraphics[width=1\textwidth]{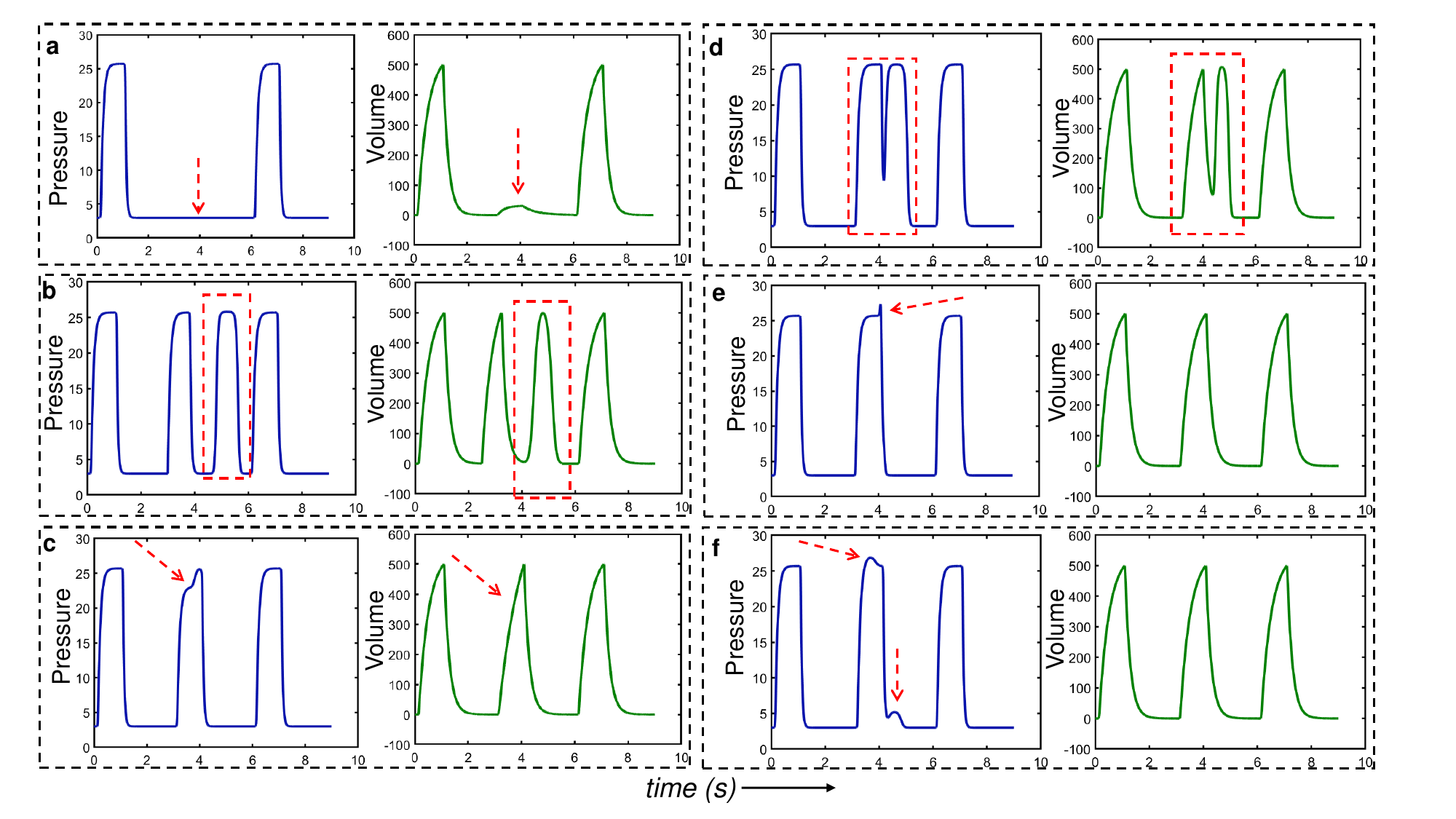}
	\caption{VDLV model produced pressure and volume VD waveforms: (a) Ineffective trigger, (b) Auto trigger, (c) Flow limited, (d) Double trigger, (e) Delayed cycling, and (f) Early cycling. Equations \ref{pressure_equation} and \ref{volume_equation} were used to simulate the pressure and volume waveforms, respectively, with parameter values shown in Tables \ref{tab:dyssynchrony} and \ref{tab:dyssynchronyV}. Note that the model variability presented here is independent of the ventilator modes.}
	\label{subplot}
\end{figure*}	

\section{Producing VD waveforms using VDLV model}
\label{section5}	
Figure~\ref{subplot} (a-f) illustrates pressure and volume waveforms corresponding to six different types of VD that are produced using the VDLV model, shown in eqs.~\ref{pressure_equation} and \ref{volume_equation}, respectively. For each VD waveform, time-dependent deformations are prominently highlighted. These deformations are produced by adjusting relevant model parameters, which are shown in Tables \ref{tab:dyssynchrony} and \ref{tab:dyssynchronyV}, respectively. In each subplot, waveforms corresponding to the normal breaths are also shown as baseline references at the beginning and end of each sequence. The pressure and volume waveforms of normal breath use only the $f_{p1}$ and $f_{v1}$ components, where the $A_{p1}$ and $A_{v1}$ parameters control the respective amplitudes. Additional parameters associated with these components change the smoothness ($\alpha_{p1}$, $\alpha_{v1}$), width ($\beta_{p1}$, $\beta_{v1}$), and the gradient of the rising ($\gamma_{p1}$, $\gamma_{v1}$) and falling ($\gamma_{p2}$, $\gamma_{v2}$) signals \cite{agrawal2024quantifiable}. Similarly, in ineffective trigger VD, where a normal breath is missing, it can be produced by setting the $A_{p1}$ parameter to zero in the pressure signal as shown in Fig. \ref{subplot} (a) and Tables \ref{tab:dyssynchrony}. A limited deformation might be observed in the volume signal, which can be produced by setting $A_{v1} = 30$, $\alpha_{v1} = 50$, $\gamma_{v1} = 100$, and $\gamma_{v2} = 200$. Since this dyssynchrony does not result in notable deformation in the pressure and volume waveforms, it is excluded from further detailed waveform analysis. 
\\
Figure~\ref{subplot} (b) illustrates auto trigger VD where an additional breath is delivered due to false non-respiratory trigger signals such as heartbeats, leaks, or fluid buildup in the circuit \cite{akoumianaki2013mechanical}. This VD is simulated using the $f_{p3}$ component, where a subsequent breath is generated following the one initiated by the $f_{p1}$ component. The amplitude and timing of auto triggered breath are controlled by the $A_{p3}$ parameter, which is usually equal to the $A_{p1}$ parameter, and the $\phi_{p3}$ parameter, respectively. Similarly, in the volume signal, the $f_{v2}$ component can be used to generate a subsequent breath where, $A_{v2}$ and $\phi_{v2}$ parameters control the amplitude and the timing of the auto triggered breath. For flow limited VD, the response of the $f_{p3}$ component is placed such that a scooping pattern that indicates flow restriction, can be produced as shown in Figure \ref{subplot} (c). The severity of flow limited VD is generally defined by the concavity that appears in the pressure waveform, which can be adjusted by the $A_{p3}$ parameter with respect to the $A_{p1}$ parameter \cite{oto2021patient,agrawal2024quantifiable}. In case of volume waveform, a reduction in rise time may be observed, which can be produced by setting $\gamma_{v1} = 100$ me. 
\\
Figure~\ref{subplot} (d) illustrates the double trigger pressure and volume VD, where two consecutive breaths are delivered without much time to exhale. In the pressure waveform, this dyssynchrony is modeled using the $f_{p1}$ and $f_{p3}$ components, where the $\phi_3$ controls the onset of the subsequent breath. This parameter is selected such that the second breath begins within 50\% of the mean inspiratory time \cite{liao2011classifying}. For the volume signal, the double trigger pattern is generated by combining $f_{v1}$ and $f_{v2}$ components, with adjustments to the phase $\phi_{v2}$ to simulate subsequent breath initiation. Parameter ranges explored for producing the corresponding volume and pressure waveform are presented in Table \ref{tab:dyssynchrony} and \ref{tab:dyssynchronyV}, respectively. Although auto-trigger and double-trigger VDs arise from different underlying mechanisms, their pressure waveforms may appear similar, particularly when a subsequent breath follows closely after the initial breath. Consequently, relying solely on airway pressure signals may not be sufficient for accurate classification~\cite{thille2007double}. Additional physiological signals, such as esophageal pressure, may be necessary to differentiate between auto trigger and double trigger VD~\cite{thille2007double }.
\\
Figure~\ref{subplot} (e) shows delayed cycling VD where a distinct deformation is observed at the end of the inspiratory phase in the pressure waveform. This deformation is produced using the $f_{p3}$ component, where the position and amplitude are controlled by the $\phi_{p3}$ and $A_{p3}$ parameters, respectively (see Table \ref{tab:dyssynchrony}).
Early cycling VD is represented in Fig.~\ref{subplot}(f) where specific deformations manifest at the onset of the inspiration and expiration part of the pressure signal. These deformations are produced using the $f_{p2}$ and $f_{p3}$ components, respectively. The positions of these components are parameterized using $\phi_{p2}$ and $\phi_{p3}$ to ensure they appear at the appropriate locations (see Table \ref{tab:dyssynchrony}). Notably, the volume waveform associated with flow limitation, delayed cycling, and early cycling closely resembles the normal volume waveform \cite{oto2021patient,de2011monitoring,de2021patient}, and thus are not separately considered for generating using GAN and cGANs.

\section{ Utilizing Generative Adversarial Networks to generate heterogeneous VD waveforms }
\label{HeterogeneityusingGAN}
				
The VDLV model can produce a range of VD waveforms with clinically relevant deformations. However, the variability of these deformations depends on the combination of parameter values. This might restrict the necessary diversity needed for effective training of ML and DL models \cite{narayan2023df}. Additionally, VD waveform can vary significantly due to the variability in individual pathophysiology, and ventilator settings \cite{hao2024adversarial}. We therefore further enhance the heterogeneity of the VDLV-produced waveforms using GANs and cGANs models. A GAN model \cite{goodfellow2020generative} consists of two neural networks: a generator ($G$) that creates synthetic signals from random noise, and a discriminator ($D$) that distinguishes between real and synthetic data. In our case, the real signal is produced using the VDLV model. We also employ conditional GAN (cGAN) \cite{mirza2014conditional}, which enhances the classic GAN framework by incorporating conditional information into the generator and discriminator. This allows for targeted data generation based on specific labels \cite{mirza2014conditional}. 
\begin{figure*}
	\centering
	\includegraphics[width=0.85\textwidth]{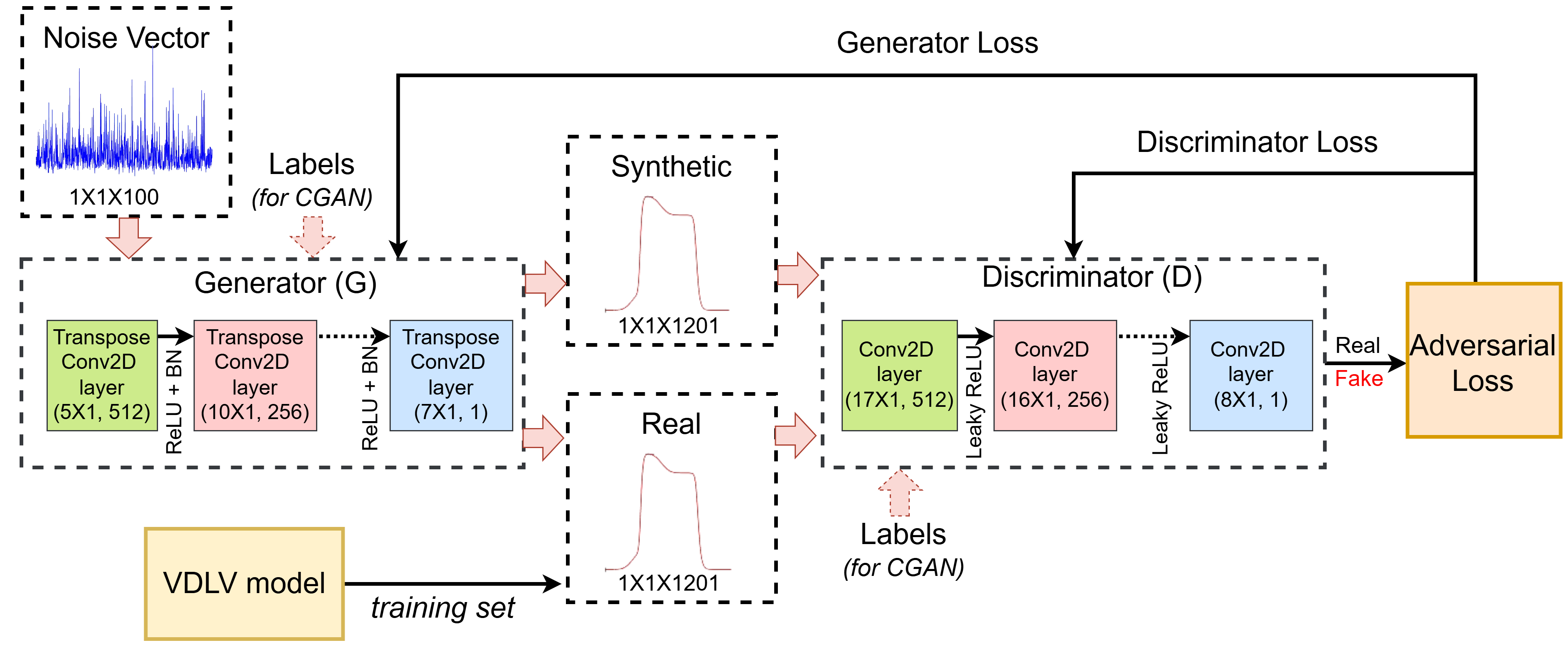}
	\caption{ Proposed architecture of the CNN-based Generative Adversarial Network (GAN) model. The model transitions into a conditional GAN (cGAN) when class-specific information is incorporated into the GAN framework. The training waveform datasets for the GAN and cGAN were produced using the VDLV model. } 
	\label{GAN}
\end{figure*}	

Our proposed GAN architecture is based on convolutional neural networks (CNNs), as illustrated in Fig.~\ref{GAN}. The generator receives a $1 \times 1 \times 100$ noise vector and produces diverse VD waveforms to mimic the training dataset by minimizing its loss. The discriminator, a CNN-based binary classifier, differentiates between real and synthetic data with a $1 \times 1 \times 1201$ input tensor. This adversarial training continues until the generator replicates the real data distribution \cite{goodfellow2020generative}. The minimax loss function that drives this process is given as:

\begin{equation}
	\label{GANeqn}
	\min_{G}\max_{D}V(D,G) = \\ \mathop{\mathbb{E}} _{x\sim p_{data}(x)}[\log D(x)]+ \mathop{\mathbb{E}} _{z\sim p_{noise}(z)}[\log (1- D(G(z)))].
\end{equation}

Here, $x$ and $z$ represent real data from the training set and the latent noise vector. The generated output is denoted as $G(z)$ while the output of the discriminator for real and generated data is represented by $D(x)$ and $D(G(z))$, respectively.
The generator aims to minimize its loss by driving $D(G(z))$ towards 1, while the discriminator seeks to maximize its loss by making $D(G(z))$ close to 0 and $D(x)$ close to 1. For the cGANs models, which utilize class labels, the loss function is adjusted as follows \cite{mirza2014conditional}:
		
\begin{equation}
	\label{cGANeqn}
	\min_{G}\max_{D}V'(D,G) = \\ \mathop{\mathbb{E}} _{x,y\sim p_{data}(x)}[\log D(x|y)]+ \mathop{\mathbb{E}} _{z\sim p_{noise}(z),y\sim p_y }[\log (1- D(G(z|y)))].
\end{equation}
		
Here, $y$ denotes the real data label vector (conditioning data). Both the generator and discriminator networks are conditioned on class label $y$, effectively guiding the data generation process \cite{mirza2014conditional}.

\subsection{Development of GAN and cGAN models } 
		
Initially, separate GANs were developed for pressure and volume waveforms for each type of VD. This is because each VD type has distinct deformations that require specialized networks to capture these features and generate realistic waveforms effectively. Each GAN generator consisted of five transposed convolutional layers, followed by batch normalization and a ReLU activation function. This architecture progressively upscaled the $1 \times 1 \times 100$ latent input vector (see Fig. \ref{GAN}). We applied batch normalization between the transposed convolutional and ReLU layers to speed up training and reduce sensitivity to initialization \cite{ioffe2015batch}. The filters sizes were $5 \times 1$, $10 \times 1$, $12 \times 1$, $5 \times 1$, and $7 \times 1$. The number of filters halved in each subsequent layer, beginning with 512 filters in the first layer. The final output is a time-series pressure or volume signal. The discriminator network consisted of five convolutional layers with filter sizes of $17 \times 1$, $16 \times 1$, $16 \times 1$, $8 \times 1$, and $8 \times 1$, employing Leaky ReLU activation function. The number of filters doubled with each layer, starting with 64. The generator had approximately 4 million learnable parameters, while the discriminator had about 2.8 million.
\\		
For cGANs, we created only two models: one for pressure and another for volume, which aimed to produce normal and all VD types waveforms. In the cGAN architecture, the generator combines a latent vector of size 100 with an embedded class label, which is reshaped and projected into a $4 \times 1 \times 786$ tensor. The discriminator concatenated the input waveform data with its corresponding class label.

\subsection{Experimental setup of GAN and cGAN models}
		
To ensure consistent training, the input data was normalized using z-score normalization, which involves subtracting the mean and dividing by the standard deviation \cite{bakkes2020machine}. We utilized the Adam optimizer for model training, with a learning rate of 0.002 and decay factors set to 0.5 and 0.999. The training was conducted using mini-batches of 256 samples over 1500 epochs. The simulations were conducted in the MATLAB R2023a environment on a system equipped with an NVIDIA GeForce RTX A4000 GPU with 16 GB of memory.
\\
We trained and evaluated the GAN model at 200, 500, 1000, 1500, and 2000 epochs to determine the optimal training duration. Generally, increasing the number of epochs enables the model to learn more complex patterns in the data, which can result in generating more realistic and plausible signals \cite{bohland2023unsupervised}. To quantitatively evaluate and compare the performance of the GAN and cGAN models in waveform synthesis, we employed a quantitative metrics that capture variability across waveform types. These metrics include Mean Absolute Error (MAE), Dynamic Time Warping (DTW), and Spectral Similarity (SS) \cite{brophy2023generative, parthasarathy2020controlled}. The MAE metric measures the average absolute difference between the generated and real waveforms and is expressed as follows:

\[
\text{MAE} = \frac{1}{l\times N} \sum_{k=1}^{N} \left| x_{\text{real(i)}}^{(k)} - x_{\text{gen(i)}}^{(k)} \right|,
\]
		
where \( l \) is the total number of elements in the signal and \( N \) is the total samples in the training and predicted dataset.
\( x_{\text{real}}(i) \) and \( x_{\text{gen}}(i) \) is the \( i \)-th values of the real and generated signals. DTW measures the similarity between two time series data by flexibly aligning sequences to minimize overall distance \cite{sakoe1978dynamic} using:
		\begin{equation}
			DTW (O,G) = \min \sqrt{\sum_{i,j}d(o_i,g_j)^2}.
		\end{equation}
		
SS quantifies similarity in the frequency domain by comparing the power spectral density of the waveforms and is defined as:
		
\begin{equation}
	SS = 1 - \frac{||P_X(f)-P_Y(f)||}{||P_X(f)||}.
\end{equation}
		
Here, \textit{generated waveforms} refer to GAN or cGAN-generated waveforms, while \textit{real waveforms} refer to the waveforms produced using the VDLV model.

		\section{Results and Discussions}
		\label{results_and_discussions}
		
\begin{figure*}[h!]
	\centering
	\includegraphics[width=1\textwidth]{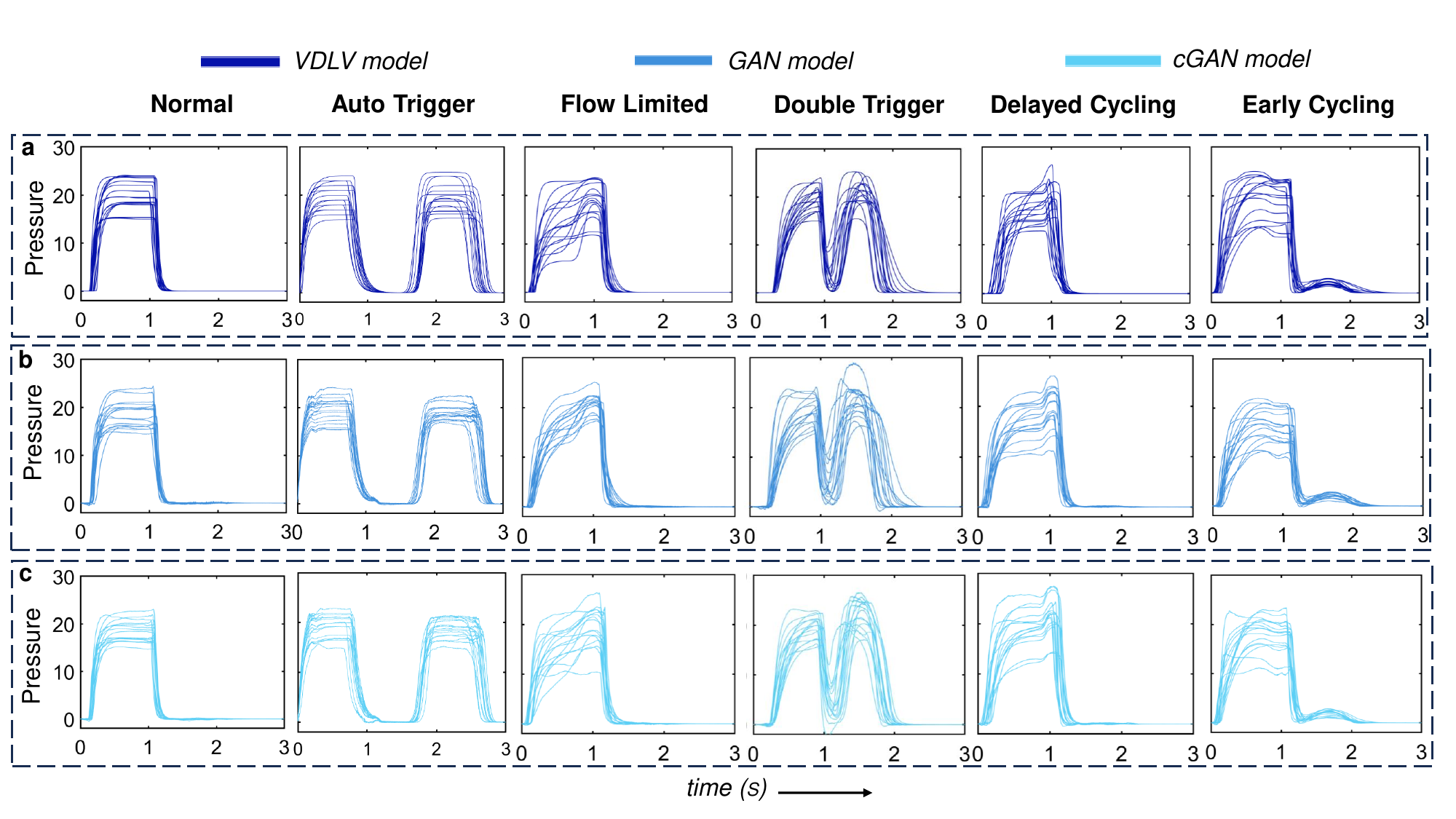}
	\caption{Sample heterogeneous pressure waveforms representing normal, auto trigger, flow limited, double trigger, delayed cycling, and early cycling waveforms produced using the (a) VDLV, (b) GAN, and (c) cGAN models.}
	\label{fig:variability}
\end{figure*}

\begin{figure*}[h!]
	\centering
	\includegraphics[width=1\textwidth]{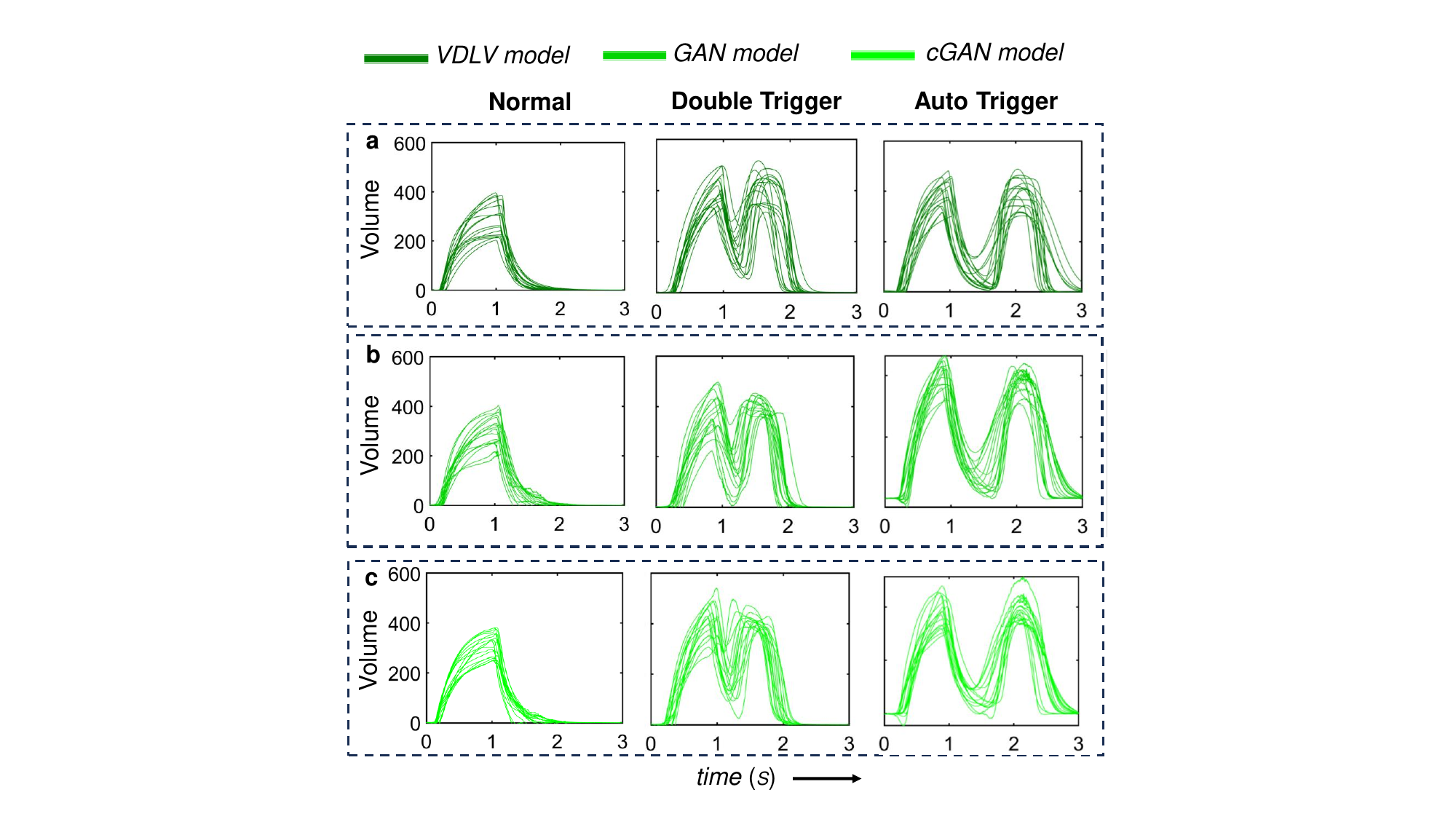}	
	\caption{Sample heterogeneous normal, double trigger, and auto trigger volume waveforms, produced using the (a) VDLV, (b) GAN, and (c) cGAN models.} 
	\label{fig:variabilityVol}
\end{figure*}

		We used the VDLV model to produce a training dataset of normal and VD-associated deformed pressure and volume waveforms for the GAN and cGAN models. Specifically, we created 1000 pressure and volume waveform signals for each breath type - normal, auto trigger, flow limited, double trigger, delayed, and early cycling. This is done by randomly selecting parameter values from predefined uniform distributions. These distributions were manually determined to produce physiologically relevant deformations, as detailed in Tables \ref{tab:dyssynchrony} and \ref{tab:dyssynchronyV}. A few sample plots for each breath type from the training datasets are shown in Figs.~\ref{fig:variability} and \ref{fig:variabilityVol}. The CP parameter was set to zero in all the training pressure waveforms to remove the constant baseline value, as it is not a VD-specific feature. This adjustment allowed the model to focus on the true waveform morphology associated with dyssynchrony events. Moreover, throughout the study, we maintained a consistent training waveform dataset to facilitate unbiased comparisons between the GAN and cGAN models.

		\begin{table}[h]
			\centering
			\caption{Parameter ranges of the VDLV model used to produce the pressure waveform dataset for training the GAN and cGAN models. The values in parentheses correspond to those used for the pressure plots displayed in Fig.~\ref{subplot}.}
			\label{tab:dyssynchrony}
			\resizebox{\textwidth}{!}{
				\begin{tabular}{|l|c|c|c|c|c|c|c|c|c|c|c|c|c|c|}
					\hline
					\textbf{Components} & 
					\multicolumn{6}{c|}{$f_{p1}(t,\theta)$} & \multicolumn{4}{c|}{$f_{p2}(t,\theta)$} & \multicolumn{4}{c|}{$f_{p3}(t,\theta)$} \\
					\hline
					\hline
					Dyssynchrony & $\alpha_{p1}$ & $\beta_{p1}$ &
					$\phi_{p1}$ & $\gamma_{p1}$ & $\gamma_{p2}$ & $A_{p1}$ & $\alpha_{p2}$ & $\beta_{p2}$ & $\phi_{p2}$ & $A_{p2}$ & $\alpha_{p3}$ & $\beta_{p3}$ & $\phi_{p3}$ & $A_{p3}$ \\
					\hline
					Normal & \makecell{20-60\\(50)} & \makecell{0.6-0.75\\(0.63)} & \makecell{-0.4} & \makecell{15-40\\(36.2)} & \makecell{10-20\\(18.17)} & \makecell{15-24\\(22.7)} & - & - & - & 0 & - & - & - & 0 \\
					\hline
					Auto Trigger & \makecell{20-60\\(50)} & \makecell{0.63-0.85\\(0.63)} & \makecell{-0.9} & \makecell{10-40\\(36.2)} &\makecell{15-40\\(18.17)} & \makecell{15-24\\(22.7)} & - & - & - & \makecell{0\\} & \makecell{2-20\\(12.2)} & \makecell{0.6-0.95\\(0.63)} & \makecell{2.2-2.8\\(2.8)} & \makecell{1-1.2$\times A_{p1}$\\(22.7)} \\
					\hline
					Flow Limited & \makecell{20-60\\(50)} & \makecell{0.5-0.65\\(0.63)} &
					\makecell{-0.4} & \makecell{15-100\\(36.2)} & \makecell{10-20\\(18.17)} & \makecell{10-22\\(20)} & - & - & - & \makecell{0\\} & \makecell{2-20\\(13.84)} & \makecell{0.97-0.99\\(0.97)} & \makecell{0.25-0.35\\(0.29)} & \makecell{0.5 to 1$\times A_{p1}$\\(2.6)} \\
					\hline
					Double Trigger & \makecell{20-60\\(50)} & \makecell{0.6-0.85\\(0.63)} &
					\makecell{-0.4} & \makecell{20-100\\(36.2)} & \makecell{10-40\\(18.17)} & \makecell{15-24\\(22.7)} & - & - & - & \makecell{0\\} & \makecell{2-20\\(10.9)} & \makecell{0.7-0.95\\(0.73)} & \makecell{1.1-1.6\\(1.52)} & \makecell{1-1.2$\times A_{p1}$\\(23)} \\
					\hline
					Delayed Cycling & \makecell{20-100\\(50)} & \makecell{0.5-0.65\\(0.63)} &
					\makecell{-0.4} & \makecell{10-100\\(36.2)} & \makecell{5-30\\(18.17)} & \makecell{10-24\\(22.7)} & - & - & - & \makecell{0\\} & \makecell{30-100\\(76.97)} & \makecell{0.98-0.99\\(0.99)} & \makecell{0.4-0.65\\(0.60)} & \makecell{0.1-0.2$\times A_{p1}$\\(3)} \\
					\hline
					Early Cycling & \makecell{20-100\\(50)} & \makecell{0.5-0.65\\(0.63)} &
					\makecell{-0.4} & \makecell{10-100\\(36.2)} & \makecell{5-30\\(18.17)} & \makecell{10-24\\(22.7)} & \makecell{5-20\\(8.2)} & \makecell{0.92-0.96\\(0.93)} & \makecell{-0.4 to 0\\(-0.4)} & \makecell{0.5-1.2\\(1.3)} & \makecell{2-10\\(9.28)} & \makecell{0.92-0.96\\(0.92)} & \makecell{1.6-1.8\\(1.6)} & \makecell{1-3\\(2.2)} \\
					\hline
					\multicolumn{15}{|c|}{$ \theta = 0.3; C.P = 0 (3)$ } \\
					\hline			
				\end{tabular}
			}
		\end{table}

\begin{figure*}[h!]
			\centering
			\begin{minipage}{0.48\textwidth}
				\centering
				\includegraphics[scale=0.25]{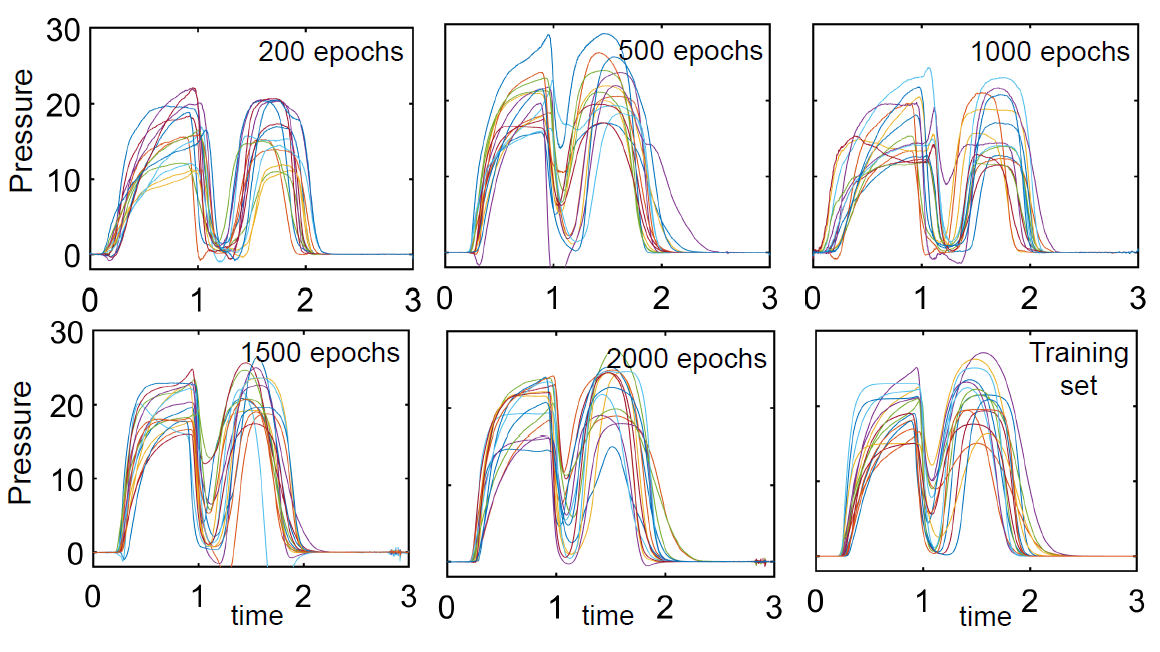}
				\caption{Few sample double trigger pressure waveforms generated using the GAN model over different epochs.}
				\label{DT_over_epochs}
			\end{minipage}%
			\hfill
			\begin{minipage}{0.48\textwidth}
				\centering
				\begin{tabular}{|c|c|c|c|p{1.2cm}|}
					\hline
					\multirow{2}{*}{\textbf{Epochs}} & \multicolumn{4}{c|}{\textbf{Generative Adversarial Network}} \\ \cline{2-5}
					& \textbf{MAE $(\downarrow)$} & \textbf{DWT $(\downarrow)$} & \textbf{SS $(\uparrow)$} & \textbf{Time (s) $(\downarrow)$} \\ \hline
					200  & 0.299 & 0.253 & 0.201 & 264  \\ \hline
					500  & 0.293 & 0.246 & 0.254 & 668  \\ \hline
					1000 & 0.276 & 0.263 & 0.408 & 1396 \\ \hline
					1500 & 0.156 & 0.090 & 0.689 & 1826 \\ \hline
					2000 & 0.106 & 0.022 & 0.731 & 2530 \\ \hline
				\end{tabular}
				\caption{Quantitative metric scores along with computational time for the  GAN model applied to double trigger pressure waveforms over different epochs.}
				\label{table_double_trigger}
			\end{minipage}
		\end{figure*}

		\begin{table}[h]
			\centering
			\caption{ Parameter ranges of the VDLV model used to produce the volume waveform dataset for training the GAN and cGAN models. The values in parentheses correspond to those used for the volume plots displayed in Fig.~\ref{subplot}.}
			\label{tab:dyssynchronyV}
			\resizebox{\textwidth}{!}{
				\begin{tabular}{|l|c|c|c|c|c|c|c|c|c|c|}
					\hline
					\textbf{Components} & 
					\multicolumn{6}{c|}{$f_{v1}(t,\theta)$} & \multicolumn{4}{c|}{$f_{v2}(t,\theta)$} \\
					\hline
					\hline
					Dyssynchrony Type & $\alpha_{v1}$ & $\beta_{v1}$ & $\gamma_{v1}$ &
					$\gamma_{v2}$ &
					$\phi_{v1}$ & $A_{v1}$ & $\alpha_{v2}$ & $\phi_{v2}$ & $\beta_{v2}$ & $A_{v2}$ \\
					\hline
					Normal & \makecell{50-100\\(96)} & \makecell{0.6-0.75\\(0.62)} & \makecell{60-200\\(150)} & \makecell{40-100\\(78)} &
					\makecell{-0.4} &
					 \makecell{200-650\\(500)} & - & - & - & 0 \\
					
					\hline
					Double Trigger & \makecell{10-100\\(96)} & \makecell{0.7-0.9\\(0.62)} 
					&
					\makecell{60-200\\(150)} & \makecell{60-100\\(78)} &
					\makecell{-0.4} & \makecell{200-650\\(500)} & \makecell{15-20\\(14.43)} & \makecell{1.2-1.7\\(1.65)} & \makecell{0.8-0.95\\(0.90)} & \makecell{1-1.2 $\times A_{v1}$\\(500)} \\
					\hline
					Auto Trigger & \makecell{10-100\\(96)} & \makecell{0.7-0.9\\(0.62)} & \makecell{60-200\\(150)} & \makecell{60-100\\(78)} &					
					\makecell{-0.9}& \makecell{200-650\\(500)} & \makecell{3-10\\(5)} & \makecell{2.2-2.7\\(2.75)} & \makecell{0.8-0.95\\(0.84)} & $A_{v1}$ \\
					\hline
					\multicolumn{11}{|c|}{$\theta = 0.3$} \\
					\hline
				\end{tabular}
			}
		\end{table}	
To identify the optimal number of training epochs for generating realistic pressure and volume waveforms, the GAN model was trained on 200, 500, 1000, 1500, and 2000 epochs. For this, we used the pressure waveform of double trigger VD, which exhibits more complex features (deformations) compared to other VD types (see Fig.~\ref{fig:variability}). Generated waveforms for each epoch are shown in Fig.~\ref{DT_over_epochs}, while quantitative results along with the training time are detailed in Table \ref{table_double_trigger}. As shown in Table \ref{table_double_trigger}, at 1500 epochs, the model achieves a substantial reduction in MAE (0.156) and DWT (0.090), along with an increase in SS (0.689), indicating that the generated double trigger pressure waveforms closely match the ground truth both numerically and structurally (see Fig. \ref{DT_over_epochs}). Compared to earlier epochs (200, 500, and 1000), these metrics show significant improvement, while the additional gains at 2000 epochs (MAE: 0.106, DWT: 0.022, SS: 0.731) are marginal and come at the cost of a much higher computation time (2530s vs. 1826s). We therefore selected 1500 epochs for further experiments.
\\		
Figures \ref{fig:variability} and \ref{fig:variabilityVol} showcase the generated pressure and volume waveforms. The first row shows the waveforms produced using the VDLV model, while the second row displays the outputs from the GAN model. Visually, a clear qualitative similarity between the GAN-generated waveforms and the training data can be seen for both pressure and volume, indicating that the GAN successfully captures the key features of the real signals. This qualitative evaluation is supported by the quantitative metrics that include MAE, DTW, and SS scores, which are listed in Table \ref{table_quantitative}. For example, lower MAE values suggest minimal average differences between generated and real waveforms, while higher SS scores indicate strong spectral similarity. Similarly, DTW scores closer to 1 reflect high alignment between the waveforms. Together, these metrics confirm that the GAN-generated waveforms closely resemble the training data, validating the GAN model ability to reproduce physiologically relevant signals.		


\begin{figure}[h]
			\centering
			\hspace{-1cm}
			\begin{subfigure}[b]{0.4\textwidth}
				\centering
				\includegraphics[scale=0.25]{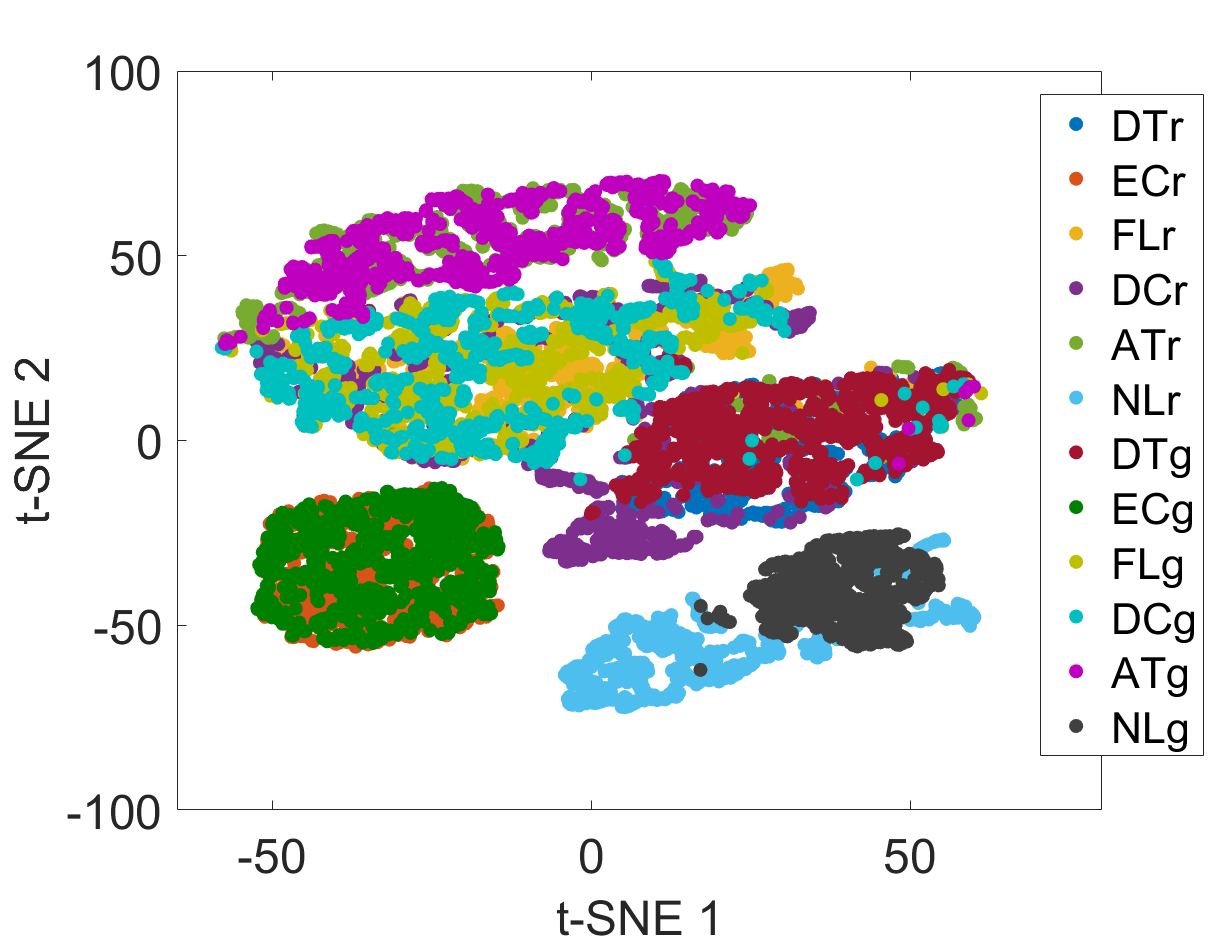}
				\caption{}
				\label{tsne_gan}
			\end{subfigure}
			\hspace{1cm}
			\begin{subfigure}[b]{0.4\textwidth}
				\centering
				\includegraphics[scale=0.25]{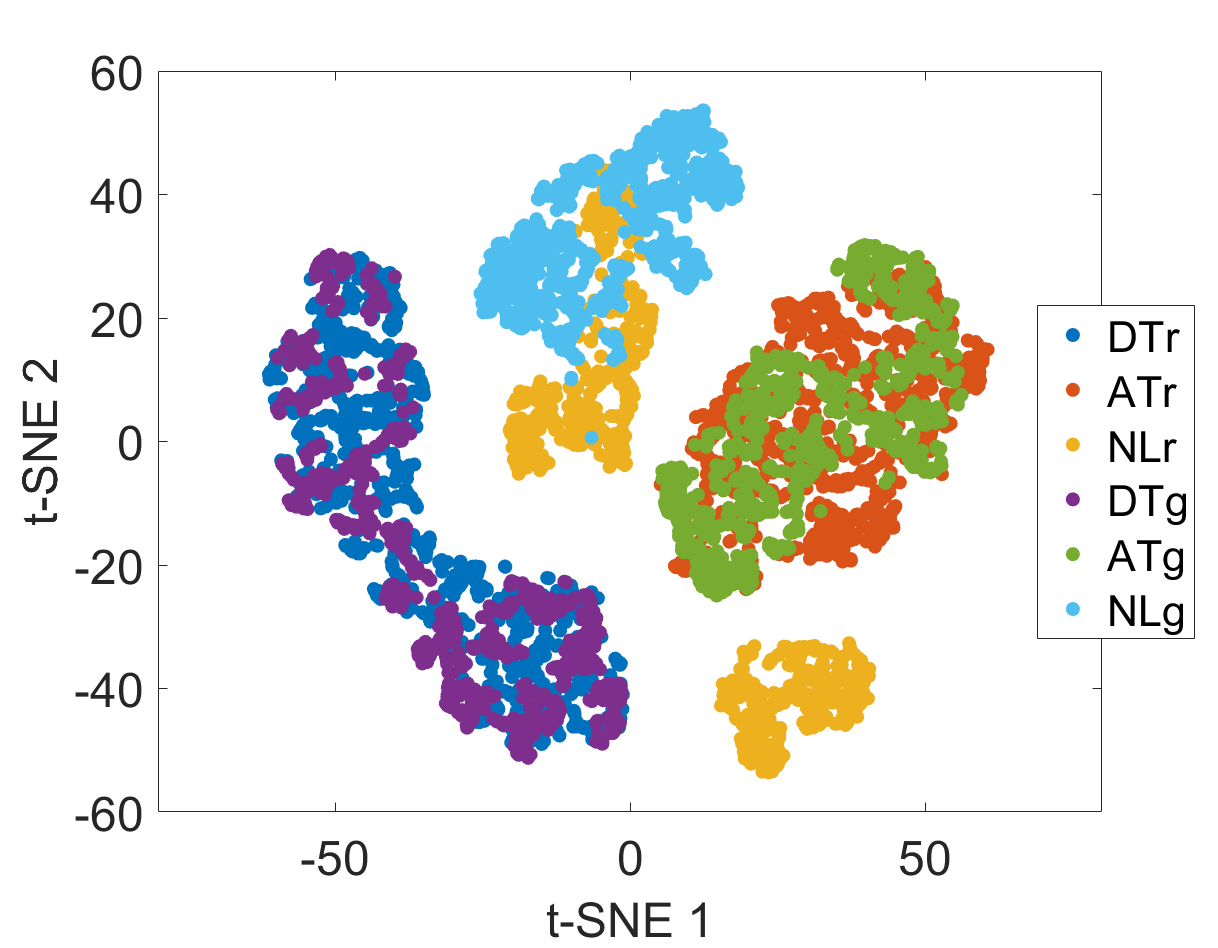}
				\caption{}
				\label{tsne_cGAN}
			\end{subfigure}
			\caption{t-SNE plots of the (a) pressure and (b) volume waveforms produced using the VDLV model \textit{(real)} and cGAN \textit{(generated)}. Here, DT, EC, FL, DC, AT, and NL denote double trigger, early cycling, flow limited, delayed cycling, auto trigger, and normal, while subscript ``r" represents the real, and subscript ``g" represents the generated.}
			\label{tsne}
		\end{figure}

The pressure and volume waveforms generated using the cGAN model are shown in the third row of Figs.~\ref{fig:variability} and \ref{fig:variabilityVol}, respectively. For the pressure waveforms, the cGAN achieved an overall MAE of 0.066, a DWT score of 0.010, and an SS score of 0.747 with a computation time of about 13920 seconds. For the volume waveform, the corresponding metrics were 0.086, 0.020, and 0.760, respectively, and the computation time was about 6660 seconds. We found that the cGAN outperformed the standard GAN for all VD waveforms as detailed in Table \ref{table_quantitative}. This improved performance can be attributed to the class-conditional learning utilized by the cGAN model, which effectively captures variations within each class. This resulted in greater heterogeneity among the cGAN generated waveforms compared to standard GAN generated waveforms \cite{bourou2024gans,cheng2020analysis}. Furthermore, while training separate GANs for each class can be resource-intensive, the cGAN optimizes this process by using a single model to handle all classes, incorporating class labels during generation and discrimination.
Figure~\ref{tsne} displays t-distributed Stochastic Neighbor Embedding (t-SNE) visualizations \cite{van2008visualizing} of the pressure and volume waveforms generated by the cGAN. The clear separation between these clusters indicates that the cGAN successfully captures distinguishing features of each VD type. This suggests that the model can generate diverse and physiologically accurate waveforms that reflect the unique characteristics of each condition. Overall, these results validate the effectiveness of combining the VDLV model with the cGAN for generating realistic, distinct pressure and volume signals for various VD waveforms.

\begin{table*}[h!]
	\centering
	\caption{Quantitative metric scores for the proposed GAN and cGAN models applied to pressure and volume VD and normal waveforms after 1500 epochs.}
	\label{table_quantitative}
	\begin{tabular}{|c|c|c|c|c|c|c|c|}
	\hline
	\multirow{2}{*}{\textbf{Signal Type}} & \multirow{2}{*}{\textbf{VD Type}} & \multicolumn{3}{c|}{\textbf{GAN}} & \multicolumn{3}{c|}{\textbf{cGAN}} \\ \cline{3-8}
	& & \textbf{MAE $ (\downarrow) $} & \textbf{DWT $ (\downarrow $)} & \textbf{SS $(\uparrow)$} & \textbf{MAE $ (\downarrow )$} & \textbf{DWT $ (\downarrow) $} & \textbf{SS $(\uparrow)$} \\ \hline
				
	\multirow{6}{*}{\rotatebox{90}{Pressure}} & Normal & 0.032 & 0.002 & 0.847 & 0.033 & 0.006 & 0.854 \\ \cline{2-2} \cline{3-8}		
	& Auto Trigger & 0.119 & 0.044 & 0.729 & 0.100 & 0.021 & 0.776 \\ \cline{2-2} \cline{3-8}
	& Flow Limited & 0.113 & 0.067 & 0.712 & 0.064 & 0.004 & 0.612 \\ \cline{2-2} \cline{3-8}	
	& Double Trigger & 0.156 & 0.090 & 0.689 & 0.077 & 0.017 & 0.779 \\ \cline{2-2} \cline{3-8}		
	& Delayed Cycling & 0.074 & 0.016 & 0.631 & 0.068 & 0.005 & 0.615 \\ \cline{2-2} \cline{3-8}
	& Early Cycling & 0.125 & 0.087 & 0.694 & 0.052 & 0.010 & 0.849 \\ \cline{2-2} \cline{3-8}
	& Overall & NA & NA & NA & 0.066 & 0.010 & 0.747 \\ \hline
	\hline
	\multirow{4}{*}{\rotatebox{90}{Volume}} & Normal & 0.044 & 0.001 & 0.801 & 0.049 & 0.007 & 0.798 \\ \cline{2-2} \cline{3-8}
	& Double Trigger & 0.167 & 0.117 & 0.694 & 0.118 & 0.032 & 0.715 \\ \cline{2-2} \cline{3-8}
	& Auto Trigger & 0.096 & 0.022 & 0.744 & 0.093 & 0.020 & 0.768 \\ \cline{2-2} \cline{3-8}
				
	& Overall & NA & NA & NA & 0.086 & 0.020 & 0.760 \\ \hline 
\end{tabular}
			\\
		\end{table*}
		
This study presents four key contributions: First, we introduced a novel synthetic waveform generator that integrates a mathematical model with GAN and cGAN models to produce pressure and volume waveforms for five types of VD. Second, we conducted a comprehensive analysis of ventilator modes and their association with various types of VD, emphasizing the key parameters contributing to the occurrence of VD. This relationship holds clinical significance, as demonstrated by Blanch et al. \cite{blanch2015asynchronies}, who observed a strong correlation between VD types and ventilator modes in a cohort of 50 patients monitored over 1700 hours. Third, we quantitatively assessed the performance of GANs and cGANs in synthesizing realistic VD waveforms. This evaluation focused on their ability to produce signals that exhibit plausible deformation patterns while maintaining sufficient diversity to represent the variability observed in clinical data accurately. Finally, we introduced a unified cGAN model capable of generating five types of VD along with normal breaths. This integrated framework simplifies waveform generation by combining multiple VD types into a single model, allowing for the simultaneous creation of diverse and well-annotated waveform signals.
\\
While the VDLV model, GAN, and cGAN frameworks demonstrated the ability to generate a diverse set of ventilator dyssynchrony (VD) waveforms, the present study has certain key limitations. First, the success of the synthetic data is dependent on the quality and diversity of the training dataset, which is produced from simulations using the VDLV model. Given that the VDLV model output is highly sensitive to its input parameters, it is important to manually explore a sufficiently broad and representative parameter range to capture realistic physiological variability. Second, our validation of heterogeneous VD waveforms involved visual inspection and quantitative metrics such as MAE, DTW, and SS scores, which may not fully capture important data quality and diversity aspects. Finally, the GANs and cGANs models may not fully capture the diversity inherent in VD waveforms. Hence, the generated waveforms need to be validated against real clinical data to confirm their clinical relevance. Future work will focus on enhancing the robustness and generalizability of the model by automating the parameter search process using systematic loops along with shape-based feature matching using local binary pattern \cite{he1990texture} or histogram of oriented gradients \cite{dalal2005histograms}.

\section{Conclusions}
\label{conclusion}
		
In conclusion, we have developed a synthetic ventilator waveform generator that combines the VDLV model with deep learning based GAN and cGAN models to create realistic waveforms for five types of VD: auto trigger, flow limited, double trigger, delayed cycling, and early cycling. The VDLV model employs functional terms that can produce physiologically relevant deformations in pressure and volume waveforms representing VD and normal waveforms. By training GAN and cGAN models on these waveforms, we significantly improved the diversity of the waveform dataset. The resulting synthetic waveforms show strong similarities to the real waveforms. This approach can help overcome the challenge posed by the limited availability of large, well-annotated VD waveform datasets.

\section*{Authorship Contribution}
S.D.D. and D.K.A. conceived and designed the research; developed the mathematical model; analyzed data and evaluated the model; S.D.D., S.D., and D.K.A interpreted results; S.D.D. prepared figures; wrote the original draft of the manuscript; S.D.D. and D.K.A. edited and revised the manuscript; approved the final version.

		\section*{Declaration of competing interest}
		The authors declare no conflicts of interest, financial or otherwise.
		
		\section*{Acknowledgements}
		This work was supported through the Seed Grant and the Institution of Eminence funding provided by the Indian Institute of Technology Bombay, India.

\section*{Source code and dataset}
The source code will be made available upon request.
		
\bibliographystyle{IEEEtran}
\bibliography{REFERENCE}
\end{document}